\documentstyle[11pt,aaspp,epsf]{article}
\newif\ifms
\msfalse

\def\refitem #1! #2! #3! #4;{\hang\noindent
    \hangindent 20pt\rm #1, \rm #2, \rm #3, \rm #4.\par}
\def\bookref{\par\noindent\hangindent 20pt}

\def\wisk#1{\ifmmode{#1}\else{$#1$}\fi}

\def\muK     {\wisk{{\rm \mu K}}}

\def\Amp     {\wisk{{\langle Q_{RMS}^2\rangle^{0.5}}}}

\def\nML     {1.55}
\def\nTB     {1.46}
\def\nTBD    {\wisk{{\nTB^{+0.39}_{-0.44}}}} 
\def\nTBDth  {\wisk{{1.25^{+0.4}_{-0.45}}}}

\def\nTBng   {\wisk{1.41^{+0.75}_{-1.17}}}  
\def\nTBcross {\wisk{1.22^{+0.42}_{-0.46}}} 

\def\SKYRMS  {GET\_SKY\_RMS}

\def\lsim   {\wisk{_<\atop^{\sim}}}
\def\gsim   {\wisk{_>\atop^{\sim}}}

\def\vs      {{\it vs.}}
\def\etal    {{\rm et al.}}

\slugcomment{astro-ph/9401015, revised, accepted by the ApJ}

\def\COBE{{\em COBE}}

\begin{document}

\title{Angular Power Spectrum of the Microwave Background Anisotropy
seen by the\\ 
\COBE\footnotemark[1] Differential Microwave Radiometer}

\footnotetext[1]{The National Aeronautics and Space Administration/Goddard
Space Flight Center (NASA/GSFC) is responsible for the design, development,
and operation of the Cosmic Background Explorer (\COBE).
Scientific guidance is provided by the \COBE\ Science Working Group.
GSFC is also responsible for the development of the analysis software
and for the production of the mission data sets.}

\author{
E.~L.~Wright\altaffilmark{2}, 
G.~F.~Smoot
\altaffilmark{3}, 
C.~L.~Bennett\altaffilmark{4}
\&
P.~M.~Lubin\altaffilmark{5}} 

\altaffiltext{2}{UCLA Astronomy Dept., Los Angeles CA 90024-1562
(I: wright@astro.ucla.edu)}
\altaffiltext{3}{LBL \& SSL, Bldg 50-351, Univ. of California, 
Berkeley CA 94720}
\altaffiltext{4}{NASA Goddard Space Flight Center, Code 685, 
Greenbelt MD 20771}
\altaffiltext{5}{UCSB Physics Dept., Santa Barbara CA 93106}

\begin{abstract}
The angular power spectrum estimator developed by 
Peebles (1973) and Hauser \& Peebles (1973) has been 
modified and applied to the 2 year maps
produced by the \COBE\ DMR.  The power
spectrum of the real sky has been compared to the power
spectra of a large number of simulated random skies
produced with noise equal to the observed noise
and primordial density fluctuation power spectra
of power law form, with $P(k) \propto k^n$.
Within the limited range of spatial scales covered by the \COBE\ DMR,
corresponding to spherical harmonic indices $3 \leq \ell \lsim 30$,
the best fitting value of the spectral index is $n = \nTBDth$
with the Harrison-Zeldovich value $n = 1$ approximately 0.5$\sigma$
below the best fit.
For $3 \leq \ell \lsim 19$, the best fit is $n = \nTBD$.
Comparing the \COBE\ DMR $\Delta T/T$ at small $\ell$ to the $\Delta T/T$
at $\ell \approx 50$ from degree scale anisotropy experiments
gives a smaller range of acceptable spectral indices which includes $n = 1$.
\end{abstract}

\section{Introduction}

The spatial power spectrum of primordial density perturbations,
$P(k)$ where $k$ is the spatial wavenumber, is a powerful tool in the
analysis of the large scale structure in the Universe.
In the first moments after the Big Bang, the horizon scale $ct$
corresponds to a current scale that is much smaller than galaxies,
so the assumption of a scale free form for $P(k)$ is natural,
which implies a power law $P(k) \propto k^n$.
Harrison (1970), Zeldovich (1972), and Peebles \& Yu (1970) all
pointed out that the absence of tiny black holes implies $n \lsim 1$,
while the large-scale homogeneity implied by the near isotropy of the
Cosmic Microwave Background Radiation (CMBR) requires $n \gsim 1$.
Thus the prediction of a Harrison-Zeldovich or $n = 1$ form for $P(k)$
by an analysis that excludes all other possibilities is an old one.
This particular scale-free power law is scale-invariant because the
perturbations in the metric (or gravitational potential) are independent
of the scale.
The inflationary scenario of Guth (1981) proposes a tremendous expansion
of the Universe (by a factor $\geq 10^{30}$)
during the inflationary epoch, which can convert
quantum mechanical fluctuations on a microscopic scale during the 
inflationary epoch into Gpc-scale structure now.  To the extent that
conditions were relatively stable during the small part of the
inflationary epoch which produced the Mpc to Gpc structures we now study,
an almost scale-invariant spectrum is produced
(Bardeen, Steinhardt \& Turner 1983).
Bond \& Efstathiou (1987) show that the expected variance of the coefficients
$a_{\ell m}$ in a spherical harmonic expansion of the CMBR temperature
given a power law power spectrum $P(k) \propto k^n$ is 
$<a_{\ell m}^2> \; \propto \Gamma[\ell+(n-1)/2] / \Gamma[\ell+(5-n)/2]$
for $\ell < 40$.
Thus a study of the angular power spectrum of the CMBR can be used to place
limits on the spectral index $n$ and test the inflationary prediction
of a spectrum close to the Harrison-Zeldovich spectrum with $n = 1$.

The angular power spectrum contains the same information as the 
angular correlation function, 
but in a form that simplifies the visualization of
fits for the spectral index $n$.  
Furthermore, the off-diagonal elements of the covariance matrix
have a smaller effect for the power spectrum than for the correlation
function.
However, with partial sky coverage the multipole estimates in the
power spectrum are correlated, and this covariance must be considered
when analyzing either the correlation function or the power spectrum.

The power spectrum of a function mapped over the entire sphere can be
derived easily from its expansion into spherical harmonics, but for a
function known only over part of the sphere this procedure fails.
Wright (1993) has modified a power spectral estimator from Peebles (1973) 
and Hauser \& Peebles (1973) that allows for partial coverage
and applied this estimator to the DMR maps
of CMBR anisotropy.
We report here on the application of these statistics to the
DMR maps based on the first two years of data (Bennett \etal\ 1994).
Monte Carlo runs have been used to 
calculate the mean and covariance of the power spectrum.
Fits to estimate \Amp\ and $n$ by maximizing the Gaussian approximation
to the likelihood of the angular power spectrum are discussed in this paper.
Since we only consider power law power spectrum fits in this paper,
we use $Q$ as a shorthand for \Amp\ or $Q_{rms-ps}$, 
which is the RMS quadrupole averaged over the whole Universe,
based on a power law fit to many multipoles.
\Amp\ should not be confused with the actual quadrupole of the 
high galactic latitude part of the sky observed from the
Sun's location within the Universe,
which is the $Q_{RMS}$ discussed by Bennett \etal\ (1992a).

\section{Estimating the Angular Power Spectrum}

Wright (1993) has discussed the modification of the Hauser-Peebles
angular power spectrum estimator for use on CMBR anisotropy maps.
We include a description of this method for completeness.
Consider a collection of spectral functions $F_{\ell m}$ which are defined
to be orthonormal in the measure $d\Omega/4\pi$.  These are the real
spherical harmonics, normalized to have an RMS value of unity for each
harmonic.  

The inner product of spatial functions $f$ and $g$ is defined as
\begin{equation}
<fg> = \int f(\Omega)g(\Omega)d\Omega/4\pi.
\end{equation}
Note that the $F_{\ell m}$'s satisfy
\begin{equation}
<F_{\ell m}F_{\ell' m'}> = \delta_\ell^{\ell'} \delta_m^{m'}.
\end{equation}

Given the temperature distribution $T(\Omega)$, define the
RMS power at each multipole $\ell$ as
\begin{equation}
T_\ell^2 = \sum_{m=-\ell}^\ell\left| <F_{\ell m}T> \right|^2
\end{equation}
The Hauser-Peebles approach to power spectra on the sphere with
non-uniform or absent coverage involved correcting for the average
density of sources: the $F_{00}$ term.  In the case of the DMR maps, we
clearly should also correct for dipole terms.

Redefine the inner product to apply to the non-uniformly covered sphere:
\begin{equation}
<fg> = {{\sum_{j=1}^N w_j f_j g_j }\over {\sum_{j=1}^N w_j }}
\end{equation}
where $j$ is an index over pixels, and $w_j$ is the weight per pixel.
In the galactic plane, $w_j = 0$.  
The galactic plane cut used in this paper excludes the 1/3 of the sky
with $|b| < 19.5^\circ$.
Outside of the galactic plane,
one can choose whether to have $w_j$ follow the map weights based on
the number of observations, $N_{obs}$.  
We have used uniform weights instead of
$N_{obs}$ weights, which increases the effect of radiometer noise
in the results, but also reduces and simplifies the correlation
between different $T_\ell$'s in the result.

Now define revised functions $G_{\ell m}$ for $\ell > 1$ given by
\begin{equation}
G_{\ell m} = F_{\ell m} 
 - {{F_{00}<F_{00}F_{\ell m}>}\over{<F_{00}F_{00}>}}
 - \sum_{m^\prime=-1}^1 {{F_{1,m^\prime}<F_{1,m^\prime}F_{\ell m}>}
                              \over{<F_{1,m^\prime}F_{1,m^\prime}>}}.
\label{Glm}
\end{equation}
These functions are orthogonal to the monopole and dipole terms in the 
region covered by the map with the specified weights.  But they are not
orthogonal to each other, nor are they normalized.
The function $G_{31}$ is substantially affected by the dipole removal,
since the galactic plane cut couples harmonics with $\Delta \ell = \pm 2$
and $\Delta m = 0$.
On the other hand $G_{13,13}$ is not much affected by the monopole
plus dipole removal but is far from normalized in the polar caps,
since most of its power is concentrated in the galactic plane.

We can now define the terms used by Hauser \& Peebles:
the normalization integral for $G_{\ell m}$
\begin{equation}
J_\ell^m = <G_{\ell m}G_{\ell m}>
\end{equation}
and the estimated spectrum
\begin{equation}
Z_\ell^m = |<G_{\ell m}T>|^2/J_\ell^m.
\end{equation}
Hauser \& Peebles recommend that the estimate to be used for the spectrum
should be the average value of the $Z_\ell^m$'s weighted by the $J_\ell^m$'s.
This quantity is
\begin{equation}
{{T_\ell^2}\over{2\ell+1}} \approx <Z_\ell^m>_m = 
{{\sum_{m=-\ell}^\ell <G_{\ell m}T>^2} \over {\sum_{m=-\ell}^\ell J_\ell^m}}.
\end{equation}

\begin{figure}[h]
\ifms
\vspace*{0.1in}
\else
\plotone{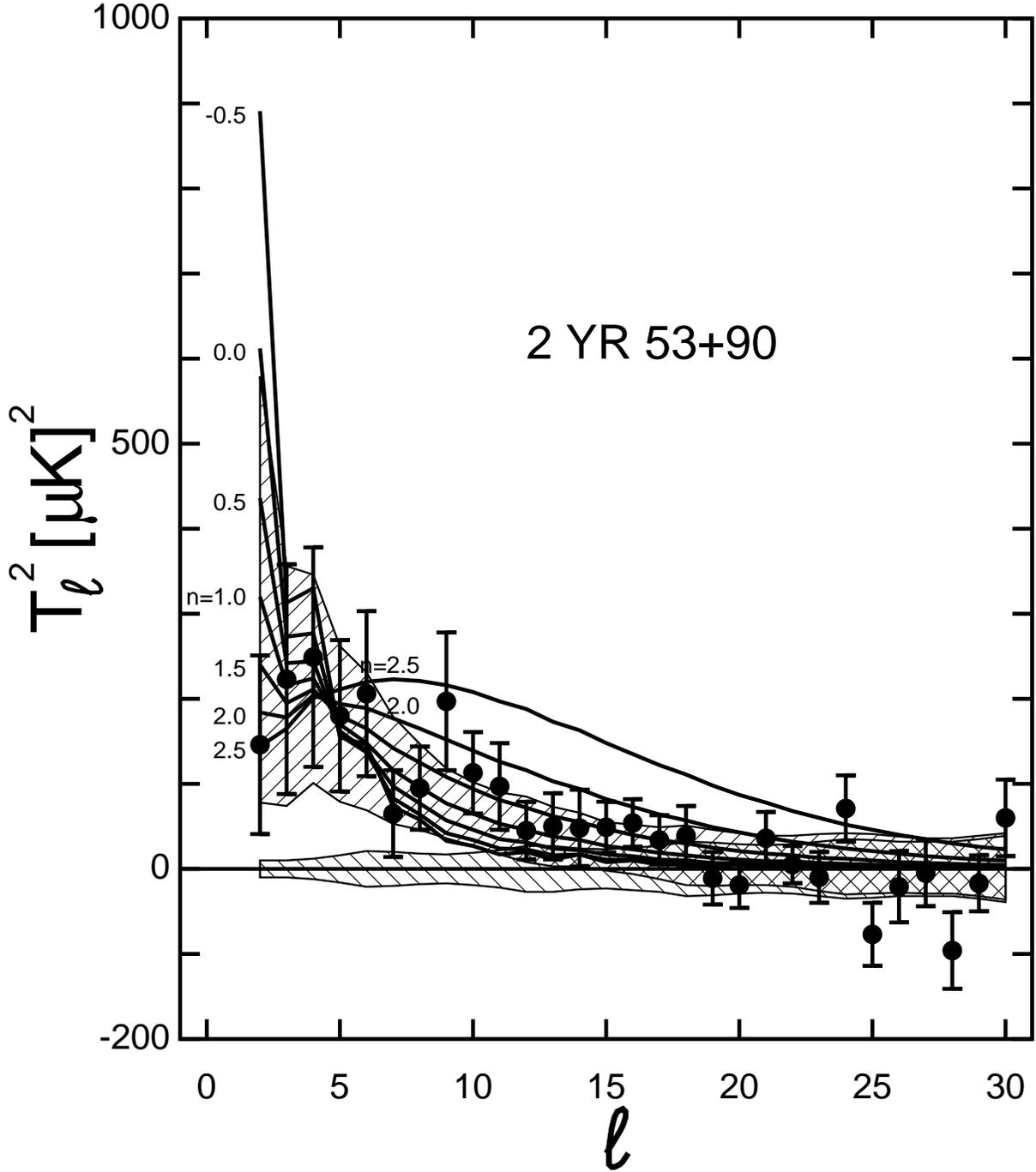}
\fi
\caption{
Power spectrum of the 2 year 53+90 DMR maps (points) compared
to the mean $\pm 1\sigma$ range of Monte Carlo spectra computed for
Harrison-Zeldovich skies with an expected Q = 17 $\mu$K.
The lower band is the $\pm 1\sigma$ range for noise only Monte Carlos.
The lines show the mean power spectra for Monte Carlo's with
$n = -0.5, 0, 0.5, 1, 1.5, 2 \;\&\; 2.5$ all normalized to 
have the same input $\ell = 4$ amplitude as the Q = 17 H-Z case.
}
\label{linear}
\end{figure}

The DMR experiment has two independent channels, $T_A$ and $T_B$, 
at each of three frequencies: 31, 53 and 90 GHz.
The sum and difference maps formed from the $A$ and $B$ channel maps
can be used to determine error associated with this estimate 
of $T_\ell^2$. 
Since $T_\ell^2$
is obtained as a sum of squares, it is necessarily positive, and is thus a
biased estimator.  The sum and difference maps can also be used to correct
this bias.  Let the sum map be $S = (T_A + T_B)/2$ while the difference is
$D = (T_A - T_B)/2$,  
where $T_A$ and $T_B$ are the maps produced by the A and B sides of the
DMR instrument.
Then an unbiased estimate of the true power spectrum 
of the sky is given by
\begin{equation}
T_\ell^2 \approx (2\ell+1)
{{\sum_{m=-\ell}^\ell\;(<G_{\ell m}S>^2 - <G_{\ell m}D>^2)} 
\over {\sum_{m=-\ell}^\ell J_\ell^m}}.
\label{psmean}
\end{equation}
These statistics evaluated for the 53+90 GHz maps with the first 2 years 
of data are shown in Figure \ref{linear}.
Assuming that both the noise map in $D$ and the cosmic plus noise map in $S$
are described by isotropic Gaussian random processes (independent of $m$),
we get an estimate for the uncertainty in $T_\ell^2$:
\begin{equation}
\sigma^2(T_\ell^2) = {{2(2\ell+1)^2\sum (J_\ell^m)^2 
\left[\left(\sum  <G_{\ell m}S>^2\right)^2 + 
\left(\sum <G_{\ell m}D>^2\right)^2 \right] } \over
{\left(\sum J_\ell^m\right)^4}}.
\label{pserror}
\end{equation}
This error estimate provides the error bars in Figure \ref{linear}.
The Monte Carlo simulations discussed below have shown that this error
estimate is correct: the mean over many simulations of the variance in
Equation \ref{pserror} agrees with the variance computed from the scatter
in the power spectra computed using Equation \ref{psmean}.
This uncertainty can easily be approximated for the case of no galactic plane
cut, and small signal to noise ratio.  In this case $J_\ell^m = 1$, and
the expected value of $<G_{\ell m}S>^2$ and
$<G_{\ell m}D>^2$ are both $\propto \sigma_1^2 / N_{tot}$,
where $\sigma_1$ is the uncertainty in a single DMR observation and
$N_{tot}$ is the total number of observations over the whole sky.
Thus the variance of $T_\ell^2$ is
\begin{equation}
\sigma^2(T_\ell^2) \propto (2\ell + 1) {{\sigma_1^4}\over{N_{tot}^2}}
\end{equation}
in this case.  For the Harrison-Zeldovich spectrum predicted by
inflation, the signal to noise ratio of $T_\ell^2$ varies like
$\ell^{-1.5}$.  Because of this rapid decrease of significance with
increasing $\ell$, we have constructed binned statistics by summing
neighboring $T_\ell^2$ into bins covering the ranges 
$\ell$ = 2, 3, 4, 5-6, 7-9, 10-13, and 14-19.  These bins are approximately
uniform in $\ln\ell$.  These binned statistics are used on plots to avoid
clutter, but the maximum likelihood fits discussed below use the unbinned
statistics.

These $T_\ell^2$'s are quadratic statistics derived from the DMR maps.
Wright \etal\ (1994) define an averaged response of a quadratic statistic 
to the spherical harmonics of a given order $\ell^\prime$.
Let 
$T^2_{\ell,\ell^\prime m^\prime}$
be the response in the $\ell^{th}$ order when the input is
the spherical harmonic $F_{\ell^\prime m^\prime}$.
The mean over $m^\prime$ of this quantity, needed to analyze isotropic 
random processes, is
\begin{equation}
T_{\ell,\ell^\prime}^2 = {{\sum_{m^\prime=-\ell^\prime}^{\ell^\prime} 
T^2_{\ell,\ell^\prime m^\prime}} \over {2\ell^\prime + 1}}
\end{equation}
Table \ref{bigtab} shows 1000 times this quantity for $\ell = 2\ldots 19$
and $\ell^\prime = 0\ldots 19$ when the galactic plane is cut
at $|b| = 19.5^\circ$.  Note the
strong coupling of orders separated by $\Delta\ell = \pm 2$ caused
by the galactic plane cut.
With no galactic plane cut, 
$T_{\ell,\ell^\prime}^2 
= \delta_\ell^{\ell^\prime}$ for $\ell \geq 2$.

\begin{figure}[h]
\ifms
\vspace*{0.1in}
\else
\plotone{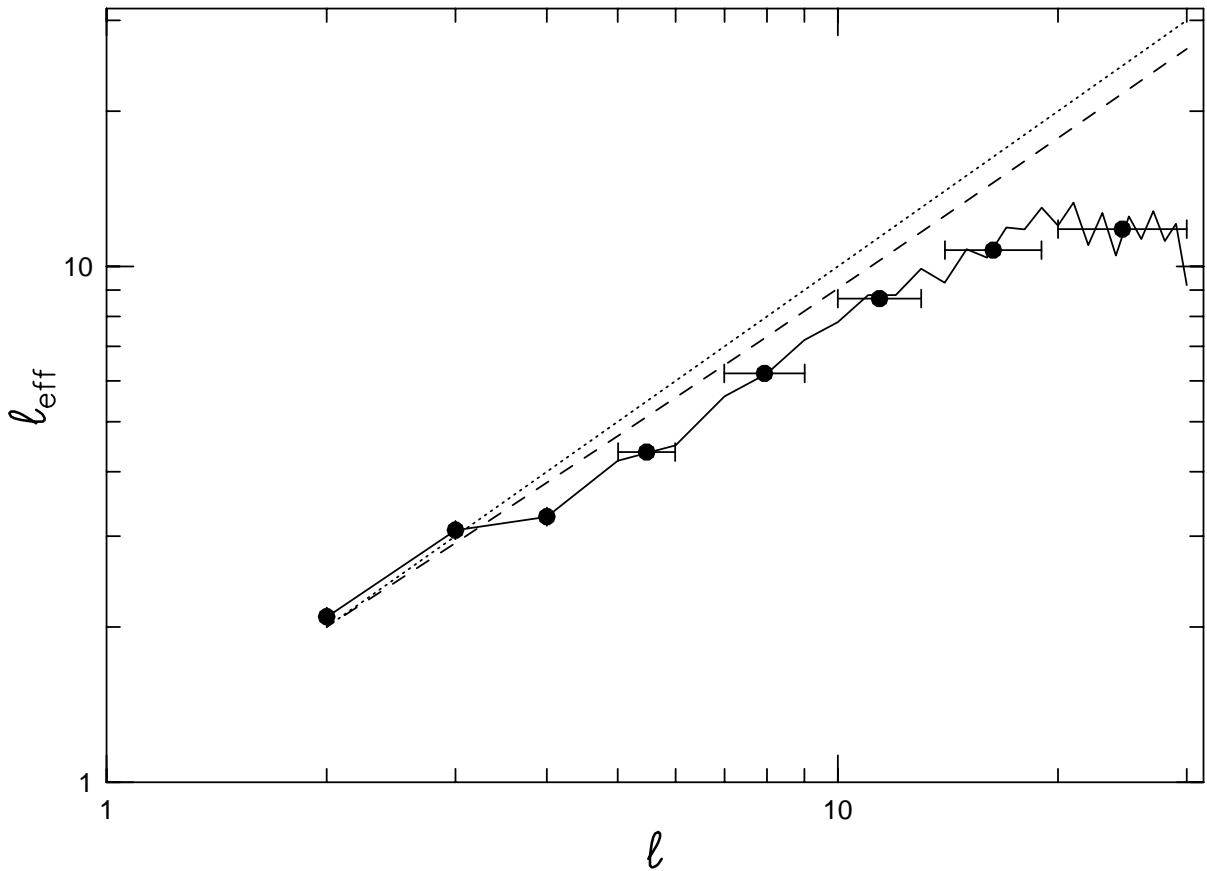}
\fi
\caption{
Effective spherical harmonic index 
$\ell_{eff}$ vs $\ell$ for the $T_\ell^2$ statistics
with a galactic plane cut of $|b| > 19.5^\circ$ (solid curve and
points for the binned statistics), and with no galactic plane cut
(dashed curve).  The dotted curve shows $\ell_{eff} = \ell$.
}
\label{elleff}
\end{figure}

The values in Table \ref{bigtab} can be used to estimate the response to
power law power spectra of primordial density perturbations
with an amplitude $Q$ and a power law index of $n$:
\begin{equation}
\overline{T_\ell^2(Q,n)} \approx Q^2 \sum_{\ell^\prime=2}^\infty 
 {{(2\ell^\prime + 1)}\over{5}} G_{\ell^\prime}^2 T_{\ell\ell^\prime}^2 
{{\Gamma[\ell^\prime+(n-1)/2] \Gamma[(9-n)/2]} \over 
{\Gamma[\ell^\prime+(5-n)/2] \Gamma[(3+n)/2]} }
\end{equation}
where $G_\ell$ is the coefficient of the Legendre polynomial expansion
of the beam given in Wright \etal\ (1994).
The effective spherical harmonic index defined by Wright \etal\ (1994),
\begin{equation}
\ell_{eff} = 2 \exp \left[ \left.{{\partial \ln \left( T_\ell^2/Q^2 \right)}
\over {\partial n}}\right|_{n=1}\right],
\end{equation}
can be evaluated either from the sum above or from the mean of Monte Carlo
simulations.  The result is that $\ell_{eff}$ is significantly smaller
than $\ell$.  The solid curve in Figure \ref{elleff} shows the relationship
for $|b| > 19.5^\circ$.
Even for the case of no galactic plane cut, $\ell_{eff}$ is smaller than
$\ell$ when $\ell > 2$, as is shown by the dashed curve in 
Figure \ref{elleff}.  
For $\ell$'s beyond the DMR beam cutoff at 
$\ell \approx 19$ the response to an $n = 1$ input spectrum is dominated
by the off-diagonal response to low $\ell$'s, so $\ell_{eff}$ saturates.
These high $\ell$ statistics are primarily sensitive to high $n$ models.

\section{Monte Carlo Simulations}

Monte Carlo simulations of the $T_\ell^2$ statistics have been done
for $n = -0.75$ to 2.75 in $P(k) \propto k^n$, and various values of $Q$.
Since the power spectrum is a quadratic function
of the sky temperatures, calculation at 3 different values of $Q$
for a given realization of the detector noise and cosmic
variance suffice to produce
the result for all values of $Q$ using quadratic interpolation.
Therefore the power spectrum of a particular Monte Carlo realization
is given by
\begin{equation}
T_\ell^2(Q,n) = a(n)Q^2 + b(n)Q + c(n)
\end{equation}
and the mean power spectrum of a set of Monte Carlo skies,
$\overline{T_\ell^2(Q,n)}$,
is given by
\begin{equation}
\overline{T_\ell^2(Q,n)} = \overline{a(n)}Q^2 + \overline{b(n)}Q + 
  \overline{c(n)}.
\end{equation}
Note that the expected values of $\overline{b(n)}$ and 
$\overline{c(n)}$ are zero,
but the actual values from a finite set of Monte Carlo simulations
will be non-zero.
The covariance matrix $C(Q,n)$ of the $T_\ell^2$ statistics is also 
determined using the Monte Carlo simulations.
Since the $T_\ell^2$ are quadratic functions
of $Q$, the covariance matrix is a quartic polynomial in $Q$.
The coefficients of the odd powers of $Q$ in this polynomial have
expected values of zero, so the covariance matrix breaks into a
noise-noise part (the coefficient of $Q^0$), a signal-noise part (the
coefficient of $Q^2$) and a signal-signal part (the coefficient of $Q^4$).
Seljak \& Bertschinger (1993) decompose the covariance matrix of the
angular correlation function in the same way.

The radiometer noise contribution to the simulated maps includes
the positive noise correlation for pixels separated by $60^\circ$
using a corrected version of the technique given in 
Wright \etal\ (1994).  The DMR maps are found by solving the matrix
equation $AT = M$ (Lineweaver \etal\ 1994), 
where $A$ is a sparse symmetric matrix, with diagonal
elements $A_{ii} = N_i$, the number of observations of the $i$'th
pixel; and off-diagonal elements $A_{ij}$ equal to minus the
number of times the $i$'th and $j$'th pixels were compared.
Wright \etal\ (1994) 
assumed that the right-hand side vector $M$ would be uncorrelated,
but it is actually anti-correlated for pixels separated by $60^\circ$.
A correct way to generate correlated noise maps is to note that
$\sigma_1^2 A^{-1}$, with $\sigma_1$ being the error in a single
sample, is the covariance matrix of the noise maps.  This implies
that noise maps can be created using $T = \sigma_1 A^{-0.5} U$,
where $U$ is a vector of uncorrelated, zero mean unit variance
Gaussian random numbers.  Even though $A$ is singular, a series
expansion of $A^{-0.5}$ converges rapidly except for the 
eigenvector corresponding to the mean of the map.  This series
is derived by writing $A = D(I+E)D$, 
with $D_{ij} = \delta_{ij}\sqrt{A_{ii}}$
and $E_{ij} = (1-\delta_{ij})A_{ij}/\sqrt{A_{ii}A_{jj}}$.
Then $A^{-0.5} \approx D^{-1}(I - 0.5 E + 0.375 E^2 - \ldots)$.
The first term gives an uncorrelated noise map, while
the second term gives a first-order correction for the 
$60^\circ$ correlation that is exactly one-half the correction 
used by Wright \etal\ (1994).
Thus we first generate a 0'th order map using uncorrelated random
numbers scaled by $N_i^{-0.5}$.
The first order correction is 1/2 of the weighted mean over the
reference ring at $60^\circ$ separation of the
0'th order map values, with the weights given by the number of times
each pixel pair is observed.
The second order correction is 3/4 of the weighted mean over the
reference ring of the first order correction.
The $m$'th order correction is $(2m-1)/(2m)$ of the weighted mean over the
reference ring of the $(m-1)$'th order correction.
A similar series approximation for the covariance matrix itself is
$A^{-1} \approx D^{-1}(I - E + E^2 - \ldots)D^{-1}$.

\section{DMR Data Selection and Power Spectrum Estimates}

The data analyzed in this paper are the maps from the first 2 years of DMR
data discussed by Bennett \etal\ (1994).
The maps are made using
pixels with cube faces oriented in galactic coordinates.
To minimize the noise, a linear combination of the 53 GHz and 90 GHz
channels is made: $0.6 T_{53}/0.931 + 0.4 T_{90}/0.815$.
The denominators in this expression convert the Rayleigh-Jeans 
differential temperatures $T_{53}$ and $T_{90}$
into thermodynamic $\Delta T$'s,
and the 60:40 weighting is used because the 53 GHz channels are the
most sensitive.
This linear combination applied to the publicly released 1 year maps
in ecliptic oriented pixels has also been analyzed.
A cross-over version of this combination, using 53A+90B and 53B+90A,
has also been analyzed.
A second linear combination used is the ``No Galaxy'' map constructed
using weights
$T_{NG} = -0.4512 T_{31} + 1.2737 T_{53} + 0.3125 T_{90}$.
This combination is calibrated in thermodynamic $\Delta T$ units,
gives zero response to the mean galactic plane and to free-free emission
(Bennett \etal\ 1992a).
The 53 GHz maps are also analyzed by themselves, using $T_{53}/0.931$
to convert to a thermodynamic $\Delta T$ scale.
Finally, the cross-power spectrum of the $53 \times 90$ GHz maps
has been found, by letting the sum map be the average of the 53 and 90
GHz maps, each converted into thermodynamic $\Delta T$'s, while the
difference map $D$ is $(53-90)/2$.
Table \ref{pstable} gives the binned power law statistics for these
four data sets.  The error bars are the square root of the diagonal
elements of the binned covariance matrix $C(Q,n)$ from the Monte Carlo runs,
evaluated at the best fit values of $Q$ and $n$, and thus include both
radiometer noise and ``cosmic variance''.  The radiometer noise for
each case is derived from the variance of the difference maps.
The ``cosmic variance'' is the error in estimating the global mean
properties of the Universe from a limited sample.  It can be estimated
from Equation \ref{pserror} and Equation \ref{psmean} in the case
where the difference map is zero, giving a limiting fractional
precision of
$\sigma(T_\ell^2)/T_\ell^2 \approx \sqrt{4\pi/[\Omega(\ell+0.5)]}$,
where $\Omega$ is the sky coverage (Scott, Srednicki \& White 1994).

\section{Maximum Likelihood Estimation}

\begin{figure}[h]
\ifms
\vspace*{0.1in}
\else
\plotone{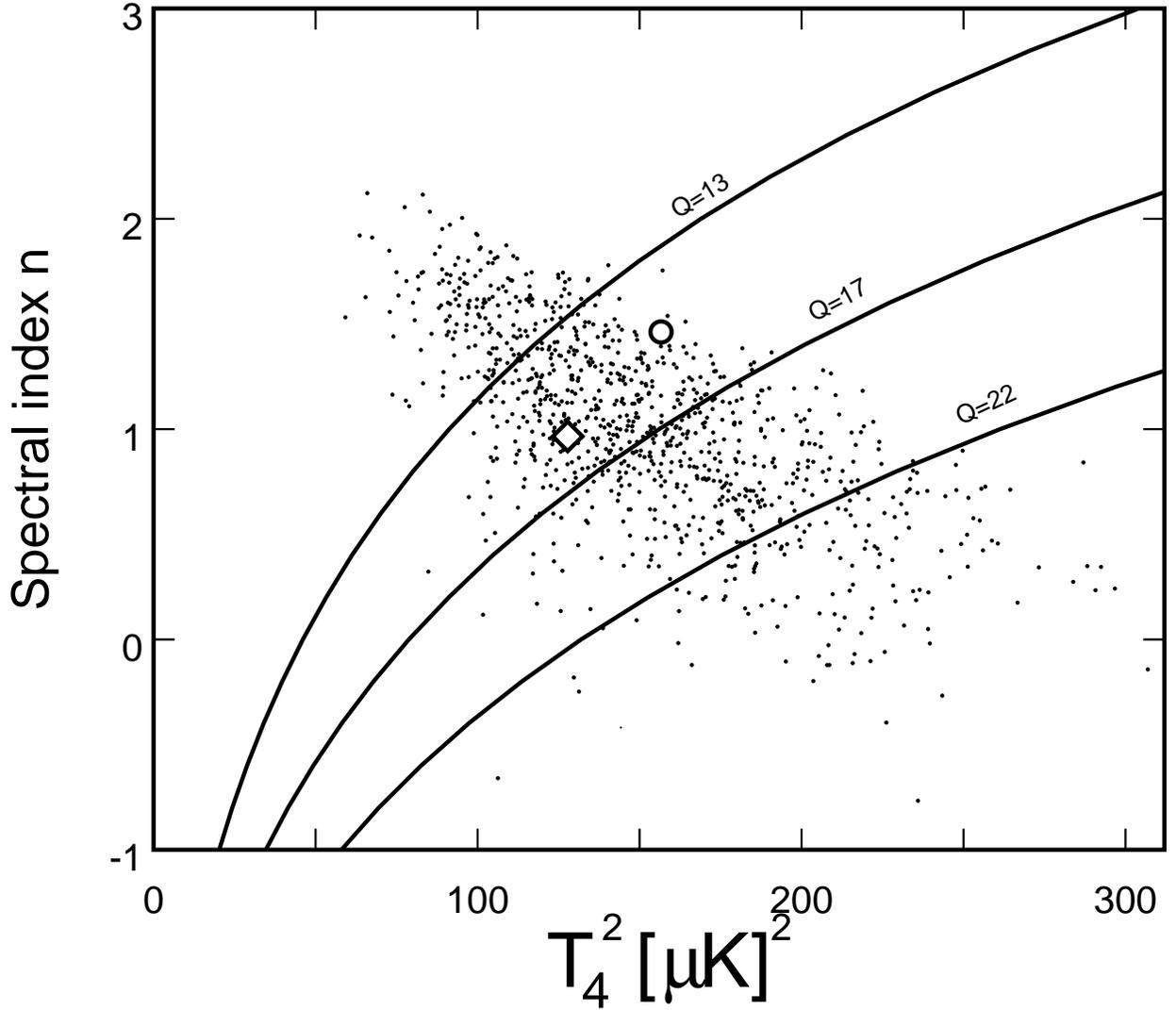}
\fi
\caption{
Scatter plot showing fitted values for the hexadecupole power
and the spectral index $n$ 
using $3 \leq \ell \leq 19$
for 1000 out of 4000 Monte Carlo skies calculated
with $\Amp = 17\;\muK$ and $n=1$. The big circle is fit to the 
power spectrum of the 2 year 53+90 GHz maps,
the big diamond is the fit to the mean of the Monte Carlo's, and the dots
are the Monte Carlo's.  The curves show combinations of $n$
and $T_4^2$ that give $Q = 13, 17 \;\&\; 22\;\muK$ (from top to bottom).
}
\label{scatter1}
\end{figure}

Given the mean power spectrum $\overline{T_\ell^2(Q,n)}$,
the covariance matrix $C(Q,n)$
and the actual power spectrum $T_\ell^2$, define
the deviation vector $e_\ell = T_\ell^2 - \overline{T_\ell^2(Q,n)}$
and the $\chi^2$ statistic $\chi^2 = e^T C^{-1} e$.
All of the fits in this paper are based on the range 
$\ell = \ell_{min}\ldots\ell_{max}$ with $\ell_{min} = 3$ and
$\ell_{max} = 19 or 30$.  
$C$ is thus a $17 \times 17$ or $28 \times 28$ matrix.  
Ignoring the quadrupole is reasonable because the galactic corrections
are largest for $\ell = 2$, and the maximum order used is set by the
DMR beam-size of 7$^\circ$ and the increased computer time required
to analyze more orders.
Since the magnitude of the covariance matrix 
gets larger rapidly when $Q$ increases 
there is a bias toward large values of $Q$ when minimizing $\chi^2$.
One can allow for this by minimizing $-2 \ln(L)$ instead of $\chi^2$,
where $L$ is the Gaussian approximation to the likelihood:
\begin{equation}
-2 \ln(L) = \chi^2 + \ln(\det(C)) + {\rm const}.
\label{gausslike}
\end{equation}
Seljak \& Bertschinger (1993) have applied this method to the correlation
function of the DMR maps.
This method has the interesting property that if the observed power spectrum
matches the model exactly then the fitted value of $Q$ is
significantly less than the true value.  At the minimum of $\chi^2$, which
is $\chi^2 = 0$ in this case, there is still a large slope in
$-2 \ln(L)$ because of the rapid variation of $\ln(\det(C))$ with $Q$.
Figure \ref{scatter1} shows this effect: the diamond symbol shows the
values of the amplitude and $n$ obtained by minimizing $-2 \ln(L)$ when the
observed power spectrum is the mean of 4000 Monte Carlo's with $n=1$ and 
$Q = 17 \;\muK$.  It is clearly biased toward
low amplitude when compared to the dots, which show fits to the individual
power spectra from the 4000 Monte Carlo's.  The big circle shows
the results of minimizing $-2 \ln(L)$ for the real sky power spectrum
from the 53+90 map with 2 years of data.

The maximum likelihood technique gives an {\it asymptotically} unbiased
determination of the amplitude $Q$ and index $n$, but only as the
observed solid angle goes to infinity.  Since we are limited to about
8~sr of sky, asymptotically unbiased means {\it biased} in practice.
In addition, the use of a Gaussian approximation for the likelihood
of our quadratic statistics can introduce additional errors.
We use our Monte Carlo simulations to calibrate our statistical
methods to avoid biased final answers.
If we maximize $L$ using the simplex method we find that
the maximum likelihood index is biased upward from the input $n$
used in the Monte Carlo's by $\approx 0.1$.
An alternative method based on finding the zero in the
finite difference $L(Q,n+0.2) - L(Q,n-0.2)$ gave a much smaller
bias but sometimes failed to converge for power spectra that were 
not well fit by a power law.

The cross power spectrum of the real sky 
based on the (53A+90A)$\times$(53B+90B) maps 
is best fit in the range $3 \leq \ell \leq 19$
by an $n = \nML$ model
when we use the simplex method to maximize the likelihood.
The improvement in $2 \ln(L)$ between the fit with $n$ forced to be 1
and the $n = \nML$ model is 2.5, which corresponds to $1.6\;\sigma$.
However, 14\% of the Monte Carlo simulations made with $n_{in} = 1$ 
and the maximum likelihood amplitude for $n = 1$ give
fitted values of $n$ that are larger than \nML, so this deviation from
a Harrison-Zeldovich spectrum is really only a ``1.06 $\sigma$'' deviation.
Similarly, 54\% of simulations made with $n_{in} = 1.5$ 
and the maximum likelihood amplitude for $n = 1.5$ had fitted indices 
higher than $\nML$, indicating that $n_{in} = 1.5$ is actually 
{\it too high} by $0.10\,\sigma$.
Using the same procedure we find that 
$n_{in} = 0.5$ is $2.07\,\sigma$ low,
$n_{in} = 2$ is $1.41\,\sigma$ high,
and $n_{in} = 2.5$ is $2.81 \,\sigma$ high.
Interpolating to find values of $n_{in}$ that deviate by -1, 0 and +1
$\sigma$ defines our quoted limits on the spectral index for
$3 \leq \ell \leq 19$: $n = \nTBD$.

With 4 years of data these limits will improve 
to $\Delta n \approx {}^{+0.32}_{-0.35}$ 
for the 53+90 maps if we assume that the maximum
likelihood $n$ remains the same.  

While waiting for this paper to be refereed, new computing facilities allowed
us to increase $\ell_{max}$ to 30.  
The increased power at $20 \leq \ell \leq 30$ expected for 
$n_{in} \approx 1.5$ is not seen in the real maps, so the fitted values of
$n$ go down.
Over the $3 \leq \ell \leq 30$ range
the fits to the cross power spectra are
$n = 1.32^{+0.39}_{-0.45}$ for (53A+90A)$\times$(53B+90B),
$n = 1.22^{+0.42}_{-0.46}$ for (53A+53B)$\times$(90A+90B), and
$n = 1.20^{+0.42}_{-0.46}$ for (53A+90B)$\times$(53B+90A).
These values have all been de-biased using the Monte Carlo simulations
as discussed above.
Each of these fits involves 4 of the 6 possible cross spectra
among the 53A, 53B, 90A and 90B maps.
Averaging these three cross spectra gives us all of the 6 possible cross
spectra.
The signal-to-noise ratio improvement from using 6 instead of 4 cross products
is quite modest, however, and is equivalent to a 22\% increase in integration
time.  Thus the adopted range for the spectral index is $n = \nTBDth$.

The spectral index from the NG maps is $n = \nTBng$,
where the large uncertainty is caused by the 
increased noise in the NG maps.
The galaxy removal process
subtracts the relatively noisy 31 GHz channels
of the DMR from a weighted sum of the quieter 53 and 90 GHz channels,
and then rescales the result to allow for the partial cancellation
of the cosmic $\Delta T$ by the subtraction.
Both the subtraction and the rescaling increase the noise, and the overall
process effectively doubles the radiometer noise.

The difference between the $n = \nTBcross$ reported here 
for $53 \times 90$ and the $n = 1.15$
reported by Smoot \etal\ (1992) is partly caused by the use of the
real beam in this paper instead of the Gaussian beam approximation
used by Smoot \etal.  The ratio of $G_{10}/G_{4}$ 
from Wright \etal\ (1994) for the real beam to the
same quantity for the Gaussian approximation is 0.92, and to compensate
for the greater suppression of $\ell = 10$ by the real beam the fitting
procedure increases $n$ by 0.2.
This increase has been partly compensated by a decrease of $n$ when
going from the 1 year to the 2 year maps.
The de-biased fit to the 1 year 
(53A+90A)$\times$(53B+90B) cross-power spectrum
for $3 \leq \ell \leq 30$ is $n = 1.69^{+0.45}_{-0.52}$.

Bennett \etal\ (1994),
using the real DMR beam instead of the Gaussian approximation, find
that the maximum of the likelihood $L(Q,n)$ occurs at
$Q = 12.4 \;\muK, n = 1.59$ from an analysis of the
cross-correlation function of the 2 year $53 \times 90$ GHz maps.
This analysis included the quadrupole, and the low observed
quadrupole leads to increased values of $n$ when it is included
in the fit.
A no quadrupole fit gives the maximum likelihood at
$n = 1.21^{+0.60}_{-0.55}$.

Smoot \etal\ (1994) give estimates of the spectral index $n$ derived
from the variation with smoothing angle of the moments of the DMR maps,
and of the genus of the DMR maps.  The determination from moments
is primarily based on the second moment, and the variation of the
second moment with smoothing angle is equivalent to
the power spectrum.  This moment method gives $n = 1.7^{+0.3}_{-0.6}$ 
when applied to the first year maps, which is quite consistent with the 
power spectrum of the first year maps.  The genus method also gives
$n = 1.7$ but $n=1$ does not give a significantly worse fit.

G\'orski \etal\ (1994) examine linear statistics that are similar
to $\langle G_{\ell m} T\rangle$.  These have the major advantage that
their distribution is exactly Gaussian, and thus the Gaussian form 
for the likelihood in Equation \ref{gausslike} is exact.  
The linear statistics used by G\'orski \etal\ define a position in
a 961 dimensional space (for $\ell \leq 30$) which is hard to visualize,
but using the exact Gaussian likelihood function
for $3 \leq \ell \leq 30$, 
G\'orski \etal\ (1994) find the 
maximum of $L(Q,n)$  occurs at $n = 1.02$ for the combined
2 year 53 GHz plus 90 GHz map.
Note that the $3 \leq \ell \leq 30$ fits in this paper still include
the small effect of the quadrupole on higher $\ell$'s due to the off-diagonal
elements in the response matrix, while those in G\'orski \etal\ (1994) are
completely independent of the quadrupole.
If Equation \ref{Glm} is modified to also subtract quadrupole terms from
the $G_{\ell m}$'s, a different modified Hauser-Peebles power spectrum is
obtained which is much more similar to the $\ell = 3-30$ analysis of
G\'orski \etal\ (1994).  
In this variant the mean power in $T_4^2$ for $n = 1$ Monte Carlo skies
goes down by 31\% while $T_4^2$ for the real sky goes up by 16\%, leading
to a higher $\ell = 4$ point that balances the high $\ell = 14-19$ bin
and reproduces the G\'orski \etal\ spectral index $n = 1.0$.
Also note that G\'orski \etal\ (1994) use a known shape and
amplitude for the noise power, 
computed from the covariance matrix of the map,
which allows them to use the ``auto-power
spectra'', while this paper just assumes that the
A and B noises are uncorrelated and can only use cross-power spectra.

Table \ref{ntable} summarizes these results and includes model-dependent
comparisons of COBE data to smaller-scale data.  The apparent large-scale
index we have used above is denoted $n_{app}$, while the model-corrected
primordial spectral index is $n_{pri}$.

Maximum likelihood fits for $Q$ with $n$ forced to be 1
over the $3 \leq \ell \leq 30$ range
give new determinations of the power spectrum amplitude:
$Q = 20.2 \pm 1.8,\; 19.2 \pm 1.3$ and $15.6 \pm 2.1 \; \muK$ 
for the 53, 53+90 and NG maps.
Comparing fits forced to $n = 0.5$ and $n = 1.5$ allows a determination
of the effective wavenumber for these amplitudes: 
$\ell_{eff} = 6.1$, 6.8 and 4.2 respectively.
The amplitudes for the 53 and 53+90 maps are higher than 
the 17 \muK\ reported earlier because the
maximum likelihood fit has emphasized the higher $\ell$'s in determining
the best fit since they have smaller cosmic variance, and
shifting to higher $\ell$'s gives a higher amplitude because the best
fit value of $n$ is greater than 1.
The values from the NG maps are statistically consistent with the
53+90 maps, but the possibility of a galactic contribution 
to $Q$ and $n$ is much reduced with the NG map.
The maps are actually more similar than the 26\% spread in best
fit $n = 1$ amplitudes would suggest: the simpler $\sigma(10^\circ)$
statistic computed using DMRSMUTH (see Wright \etal\ 1994)
in $|b| > 30^\circ$ is 31.9, 32.6 and 31.4 \muK\ for the 
53, 53+90 and NG maps respectively,
a spread of only 4\%; while the \SKYRMS\ program in
$|b| > 20^\circ$ gives $\sigma(10^\circ) = 31.3$, 29.1 and 30.7 \muK,
a spread of only 7\%.
Thus most of the difference
in the best fit amplitudes is caused by the shift of the weights to
higher $\ell$'s.

\section{Comparison with Degree Scale Experiments}

Several groups have reported statistically significant signals from
$\Delta T$ experiments with beam sizes and chopper throws close to 1$^\circ$.
These results are usually reported as limits on the amplitude of a
Gaussian correlation function,
\begin{equation}
C_g(\theta) = C_g(0)\exp(-0.5\theta^2/\theta_c^2).
\end{equation}
We have calculated the conversion from the reported limits on Gaussian
$C_g(0)$ to limits on power law power spectra as follows:
first, given the size of a Gaussian approximation to the experiment beam,
$\sigma_B = $FWHM$/\sqrt{8\ln 2}$, find the beam smoothed Gaussian
correlation function:
\begin{equation}
C_{g,sm}(\theta) = C_g(0) {{\theta_c^2}\over{\theta_c^2+2\sigma_B^2}}
   \exp[-0.5\theta^2/(\theta_c^2+2\sigma_B^2)].
\end{equation}
Then the single subtracted, double subtracted, 
triple subtracted (Python)
or square pattern $({}^+_-{}^-_+)$ double subtracted (WD2) 
temperature difference is found from
\begin{eqnarray}
{\rm var}(\Delta T_{SS}) & = & 2 (C_{g,sm}(0) - C_{g,sm}(\theta)) \nonumber \\
{\rm var}(\Delta T_{DS}) & = & 1.5 C_{g,sm}(0) - 2 C_{g,sm}(\theta) + 0.5
C_{g,sm}(2\theta) \nonumber \\
{\rm var}(\Delta T_{TS}) & = & 1.25 C_{g,sm}(0) - 1.875 C_{g,sm}(\theta) 
+ 0.75 C_{g,sm}(2\theta) - 0.125 C_{g,sm}(3\theta)\nonumber \\
{\rm var}(\Delta T_{SQ}) & = & C_{g,sm}(0) - 2 C_{g,sm}(\theta) 
+  C_{g,sm}(\sqrt{2}\theta)
\label{chopDT}
\end{eqnarray}
where $\theta$ is the chopper throw.
The same temperature differences are then estimated for power law power 
spectra with $Q = 17 \;\muK$ and $n = 0.5, 1,$ and 1.5 using
the expression for the beam smoothed power law correlation function
\begin{eqnarray}
C_{n,sm}(\theta) & = & Q^2
{{\Gamma[(9-n)/2]} \over {\Gamma[(3+n)/2]}} \nonumber \\
& \times &
\sum_{\ell=2}^\infty {{(2\ell+1)}\over{5}} P_\ell(\cos\theta)
\exp[-\ell(\ell+1)\sigma_B^2] 
{{\Gamma[\ell+(n-1)/2]}\over {\Gamma[\ell+(5-n)/2]}}.
\label{cnsm}
\end{eqnarray}
With $Q$ held fixed, we can define the effective spherical harmonic
order for a given experiment using
\begin{equation}
\ell_{eff} = 2 \times {{ {\rm var}(\Delta T[n=1.5])}\over
{{\rm var}(\Delta T[n=0.5]}}
\end{equation}
where ${\rm var}(\Delta T)$ is one of the four expressions in 
Equation \ref{chopDT}
with $C_{g,sm}$ replaced by $C_{n,sm}$ and the choice of
which expression for ${\rm var}(\Delta T)$ to use depends on the
chopping strategy used in each experiment.
The ratio of the observed ${\rm var}(\Delta T)$ derived from $C_{g,sm}$
in Equation \ref{chopDT} to ${\rm var}(\Delta T[n=1])$ with 
$Q = 17\;\muK$ then defines the $y$-axis coordinate in Figure
\ref{logngpower}, while $\ell_{eff}$ gives the $x$-axis coordinate.

Ganga \etal\ (1993, 1994) have 
have estimated the power spectral
parameters $Q$ and $n$ using data from the  
Far Infra-Red Survey (FIRS), a balloon-borne survey experiment with a 
$3.8^\circ$ beam. 
They obtain $n = 1.0^{+1.1}_{-1.0}$
which is consistent with the value in this paper.  From the slope of their
likelihood contours in the
$Q-n$ plane, we derive an $\ell_{eff} = 6.6$ for their $n = 1$ amplitude
determination, giving the FIRS point on Figure \ref{logngpower}.

\section{Conclusions}

The Hauser-Peebles method of analyzing angular power spectra has been
applied to the DMR maps.  
While the best fit to the observed power
spectrum has $n = \nTBDth$, this deviation from the $n = 1$ case is not 
statistically significant.
To obtain a more accurate determination of $n$ we need to compare the
\COBE\ DMR amplitude with ground-based and balloon-based experiments
at smaller angular scales, which are sensitive to higher $\ell$'s than
\COBE.

\begin{figure}[h]
\ifms
\vspace*{0.1in}
\else
\plotone{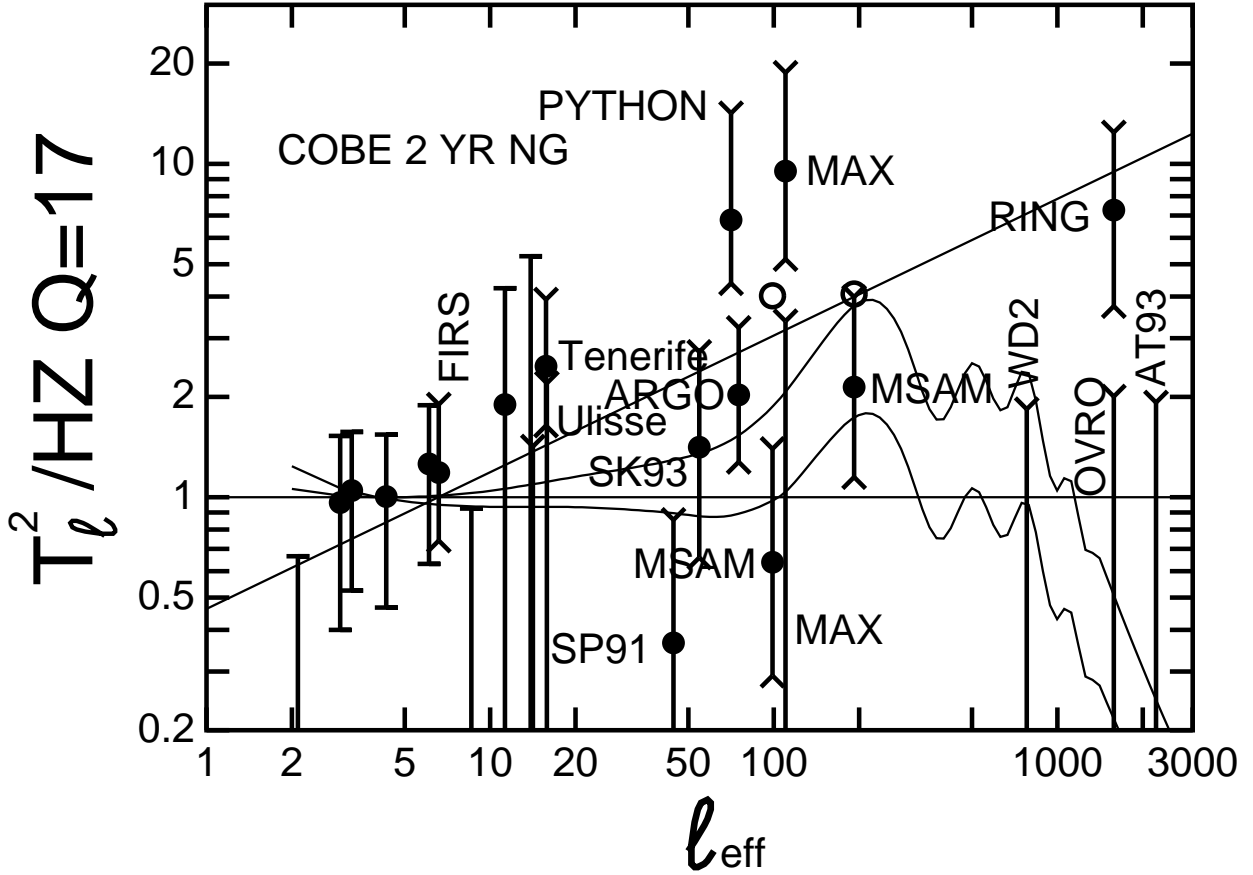}
\fi
\caption{
Power spectra normalized to the mean of 17 $\mu$K Harrison-Zeldovich
Monte Carlo skies.  \COBE\ data points from the 2 year NG DMR maps.
Models shown as thin curves:
$n = 1$, $Q = 17$ \muK\ is the horizontal line,
the best fit $n = 1.4$ power law is the slanted line,
\& tilted CDM including the effects of gravitational waves
with the long dashed curve showing $n = 0.96$ 
(predicted by $\phi^4$ chaotic inflation),
and the short dashed curve showing $n = 0.85$ where the 
tensor and scalar quadrupoles are equal (Crittenden \etal\ 1993).
Points with ``bent'' ends on their error bars are from other experiments:
FIRS (Ganga \etal\ 1993),
(from left to right)
ULISSE (de Bernardis \etal\ 1992), 
Tenerife (Watson \etal\ 1992 and Hancock \etal\ 1994),
the South Pole (Schuster \etal\ 1993),
Saskatoon (Wollack \etal 1993), 
the Python experiment (Dragovan \etal\ 1994), 
ARGO (de Bernardis \etal\ 1994),
MSAM single subtracted (Cheng \etal\ 1994),
MAX (Gunderson \etal\ (1993) and Meinhold \etal\ (1993)),
MSAM double subtracted, 
White dish second harmonic (Tucker \etal\ 1993),
OVRO (Readhead \etal\ 1989), OVRO RING (Myers \etal\ 1993),
and the Australia Telescope (Subrahmayan \etal\ 1993).
The open circles above the MSAM points show the effects of not removing
sources.
}
\label{logngpower}
\end{figure}

The ULISSE and Tenerife experiments (Watson \etal\ 1992)
with beam sizes near $6^\circ$
and chopper throws of $6-8^\circ$ give upper limits in the
$\ell_{eff} \approx 15$ range which support $n < 1.5$.
The new Tenerife results (Hancock \etal\ 1994) at the same angular
scale as Watson \etal\ give a central value that is slightly
above the earlier upper limit.  Both are plotted in 
Figure \ref{logngpower}.

The calculations of Kamionkowski \& Spergel (1994) suggest that for open
Universes with $\Omega \approx 0.1$ the power at low $\ell$'s will be
depressed relative to the $n = 1$ flat Universe prediction.  
This prediction is
consistent with the data presented here, but fluctuations due to cosmic
variance at low $\ell$'s are as large as the difference between the
open Universe model and the scale-invariant flat Universe $n=1$ model.

The string model prediction given by Bennett, Stebbins \&
Boucher (1992b) also has lower power at small $\ell$'s, and is
thus consistent with the \COBE\ angular power spectrum, but a
cutoff at higher $\ell$ is needed.  
Reionization of the Universe at redshift
$z$ will hide structures on scales smaller than $60^\circ/\sqrt{1+z}$ and
provide the needed cutoff, but $z \gsim 100$ is required to have a 
substantial optical depth with the baryon abundance derived from
Big Bang nucleosynthesis (Walker \etal\ 1991).
But reionization will not ``smear out'' the 
edges produced by strings seen at smaller redshifts.  
Thus, unlike most scientific models which can only be falsified, 
the string model can be {\em verified} by finding ``the edge'', 
which will remain infinitely sharp even with reionization.
The sharp edges in the
$\Delta T$ maps produced by nearby strings limits the slope of the cutoff
to $\ell^{-1}$ relative to an $n = 1$ spectrum.
Graham \etal\ (1994) find that the SP91 data is
significantly non-Gaussian, which suggests that an edge may have been
found.  If true, this would increase the discrepancy among the 
degree-scale experiments, since the presence of an edge would increase the
variance, but SP91 has the smallest variance of the four degree-scale
experiments.

This expected increase in the variance due to non-Gaussian features is
clearly present in the 20 GHz OVRO and RING experiments which have the 
same angular scale.  The RING experiment covered a larger region, and was
contaminated by discrete sources whose existence was verified by the VLA.
The 170 GHz MSAM experiment also saw what appeared to be discrete sources,
and these were not included in the analysis.
There is no sensitive, higher angular resolution telescope to verify
that the large deflections seen by MSAM are indeed point sources.
Thus it is possible that the large deflections are true cosmic 
$\Delta T$'s.
The open circles above the MSAM data points in Figure \ref{logngpower}
show the increased power that results if these data are not excluded 
in the analysis.

The bulk flow data of Bertschinger \etal\ (1990) 
(at $\ell_{eff} \approx 10^2$)
require $n \approx 1$ to agree with \COBE\ (Wright \etal\ 1992), 
while the larger bulk flow 
on larger scales seen by Lauer \& Postman (1992) 
requires $n \approx 2.9$ to agree with \COBE, if we assume that the
reported bulk flow represents the RMS velocity on this scale.

The experiments at $\approx 1^\circ$  scale offer the possibility of
a better determination of the primordial power spectrum index $n$,
but the model-dependent effects of the wing of the Doppler peak
at $\ell \approx 200$ must be allowed for.  
Even in the large angle region $\ell < 30$ small model-dependent
corrections must be made.  In Figure \ref{logngpower}, the upper 
Cold Dark Matter (CDM)
curve has a primordial spectral index $n_{pri} = 0.96$, but an apparent 
index $n_{app}$ = 1.1.
Since the spectral index \nTBDth\ found in this paper is an apparent
index, the \COBE\ power spectrum is consistent with the prediction
from inflation and CDM or Mixed Dark Matter models that $n_{app} \approx 1.1$.
On the other hand, vacuum-dominated models such as the
Holtzman (1989) model with
$\Omega_B = 0.02$, $h = 1$, $\Omega_{CDM} = 0.18$, and $\Omega_{vac} =
0.8$ will give $n_{app} = 0.9$ for $n_{pri} = 1.0$
(Kofman \& Starobinski 1985), which deviates by
slightly less than $1\sigma$ from the \COBE\ value.
We have found the values of the primordial spectral index $n_{pri}$ that will
connect the \COBE\ NG amplitude and $\ell_{eff}$ found earlier
with the degree-scale experiments using the scalar transfer function from 
Crittenden \etal\ (1993).
We have ignored the tensor transfer function because the current
accuracy in determining $n$ is not sufficient to fix the
tensor to scalar ratio, and because the excess quadrupole predicted
by the tensor transfer function is not seen in the \COBE\ power
spectrum.
The South Pole experiment of Schuster \etal\ (1993) at $\ell \approx 44$
requires $n_{pri} \approx 0.48 \pm 0.34$ to agree with \COBE.
The Saskatoon experiment of Wollack \etal\ (1993) at $\ell \approx 55$
requires $n_{pri} \approx 1.04 \pm 0.29$ to agree with \COBE.
The PYTHON experiment of Dragovan \etal\ (1994) at $\ell \approx 71$
requires $n_{pri} \approx 1.58 \pm 0.22$ to match \COBE.
The ARGO experiment of de Bernardis \etal\ (1994) at $\ell \approx 75$
requires $n_{pri} \approx 1.10 \pm 0.16$ to match \COBE.
The weighted mean of these values is $n_{pri} = 1.15 \pm 0.11$.
Unfortunately $\chi^2 = 8.0$ with 3 degrees of freedom when comparing
these four values of $n_{pri}$ with this weighted mean, indicating
that these four experiments are mutually inconsistent.
If we allow for this discrepancy by scaling the error on $n_{pri}$, we get
a value $n_{pri} = 1.15 \pm 0.2 \pm 0.1$ from this comparison of \COBE\ with 
the degree-scale experiments, where the second error bar is contribution
of the uncertainty of the \COBE\ NG amplitude to $n_{pri}$.
Thus this comparison of \COBE\ with degree-scale experiments
gives a more precise value the primordial spectral index that is still
consistent with inflation.
With more data from COBE (4 years are recorded) the large angular scale
amplitude will become more and more certain.  The $\ell_{eff}$ 
associated with this amplitude will shift to larger values $\approx 10$.
Reliable, consistent determinations of $\delta T$ on scales 
$\ell_{eff} \approx 50$ will be needed to compare with the large-scale
$\Delta T$.  
With only two years of data, the \COBE\ DMR large scale amplitude has 
relative errors that are two times smaller than the errors of the 
current degree-scale experiments.
Thus the degree-scale experiments need to be extended to a sky coverage
that is 10 times higher than their current coverage to match
the expected \COBE\ uncertainty with four years of data, or else achieve
an equivalent increase in accuracy by reduced noise or systematic errors.

\bigskip

We gratefully acknowledge the many people who made this paper possible:
the NASA Office of Space Sciences, the {\it COBE} flight operations team, 
and all of those who helped process and analyze the data.
In particular we thank Tony Banday, Krys G\'orski,
Gary Hinshaw, Charlie Lineweaver, Mike Hauser, Mike Janssen,
Steve Meyer and Rai Weiss for useful comments on the manuscript.

\clearpage

\begin{table*}
\plotone{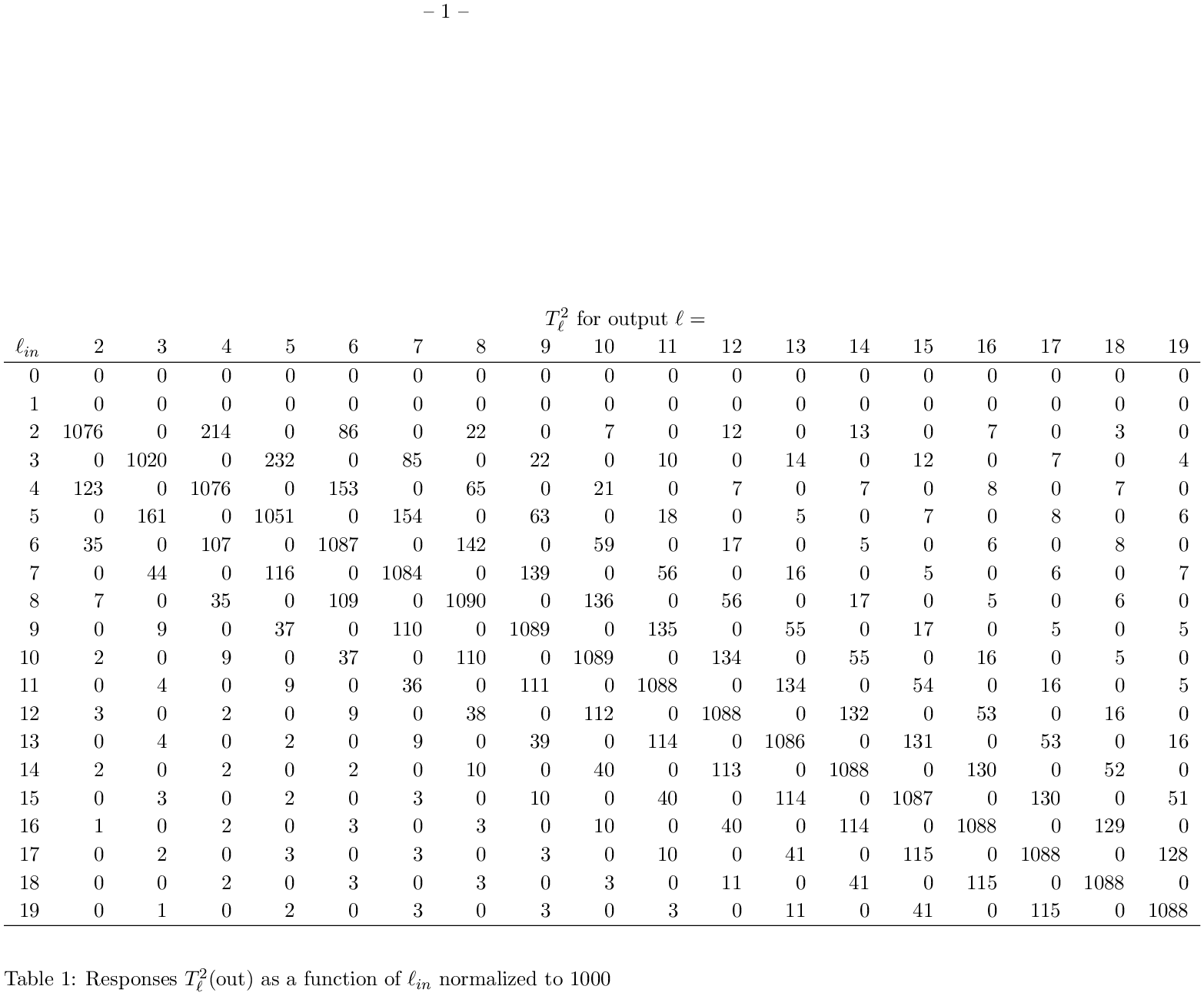}
\caption{}
\label{bigtab}
\end{table*}

\clearpage

\begin{table*}
\begin{center}
\begin{tabular}{ccrlrlrlrlrl}
$\ell$ & $\ell_{eff}$ & 
\multicolumn{2}{c}{2 YR 53} 
& \multicolumn{2}{c}{2 YR 53+90} 
& \multicolumn{2}{c}{2 YR 53$\times$90}
& \multicolumn{2}{c}{2 YR NG}
& \multicolumn{2}{c}{1 YR 53+90}\\
\tableline
 2    & 2.1  & 0.59 & $\pm 0.47$ & 0.44 & $\pm 0.61$ 
& 0.45  & $\pm 0.58$ & 0.17  & $\pm 0.49$ & 0.58 & $\pm 0.42$ \\
 3    & 3.1  & 1.06 & $\pm 0.52$ & 1.04 & $\pm 0.60$ 
& 1.01  & $\pm 0.56$ & 0.96  & $\pm 0.56$ & 0.90 & $\pm 0.50$ \\
 4    & 3.3  & 1.16 & $\pm 0.45$ & 1.11 & $\pm 0.52$ 
& 1.12  & $\pm 0.48$ & 1.05  & $\pm 0.52$ & 1.14 & $\pm 0.46$ \\
5-6   & 4.4  & 1.23 & $\pm 0.40$ & 1.21 & $\pm 0.40$ 
& 1.22  & $\pm 0.36$ & 1.00  & $\pm 0.54$ & 1.06 & $\pm 0.42$ \\
7-9   & 6.2  & 1.03 & $\pm 0.43$ & 1.23 & $\pm 0.37$ 
& 1.19  & $\pm 0.33$ & 1.26  & $\pm 0.63$ & 1.19 & $\pm 0.46$ \\
10-13 & 8.7  & 1.31 & $\pm 0.58$ & 1.51 & $\pm 0.44$ 
& 1.15  & $\pm 0.41$ & -0.22 & $\pm 1.12$ & 1.41 & $\pm 0.71$ \\
14-19 & 10.7 & 2.61 & $\pm 0.91$ & 1.60 & $\pm 0.67$ 
& 1.69  & $\pm 0.65$ & 1.90  & $\pm 2.34$ & 2.34 & $\pm 1.27$ \\
20-30 & 11.8 & 3.26 & $\pm 2.60$ & -1.14 & $\pm 1.98$ 
& 0.58  & $\pm 1.94$ & -2.54  & $\pm 7.83$ & -0.19 & $\pm 3.98$ \\
\hline
\end{tabular}
\end{center}
\caption{Ratio of the binned power spectrum from Equation 9 
to a $Q = 17\;\muK$, $n = 1$ model.}
\label{pstable}
\end{table*}

\begin{table*}
\begin{center}
\begin{tabular}{rrrrrrrrr}
$\ell=$ & 2 & 3 & 4 & 5-6 & 7-9 & 10-13 & 14-19 & 20-30 \\
\tableline
2 yr NG  &    55 &   208 &   234 &   319 &   365 &   -45 &   251 &  -161 \\
Best fit &   202 &   159 &   170 &   272 &   286 &   229 &   167 &    85 \\
Q=17,n=1 &   316 &   216 &   225 &   318 &   290 &   202 &   132 &    63 \\
                            \\
Noise only
         &  1374 &     1 &   253 &   181 &   -42 &   -28 &   -12 &    -3 \\
         &       &  1660 &    11 &   745 &   213 &    -1 &   238 &   311 \\
Signal \& Noise 
         & 24871 &       &  2164 &   669 &    30 &   133 &   171 &  -186 \\
         &   364 & 14803 &       & 10865 &  2018 &   139 &   226 & -1782 \\
         &  3611 &  -166 & 13599 &       & 18437 &  2307 &   -28 & -1230 \\
         &  1717 &  3922 &  1214 & 29146 &       & 42218 &  5603 &   264 \\
         &   828 &  1426 &   720 &  3907 & 33123 &       & 87028 & 10292 \\
         &  2055 &  -187 &   384 &  1078 &  3798 & 53633 &       & 241805 \\
         &  1128 &   776 &  -354 &   949 &  -241 &  7243 & 95862 &       \\
         & -351  &  -211 &   430 & -2326 & -1799 &   229 &  9420 & 245629 \\
\hline
\end{tabular}
\end{center}
\caption{Binned Hauser-Peebles power spectrum of the 2 year NG maps,
the best fit $n=1.4$ model, the nominal $Q=17, n=1$ model, all in $\muK^2$;
the upper triangle of the covariance matrix from noise-only Monte Carlo
runs, and the lower triangle of the covariance matrix from the best fit
Monte Carlo runs in $\muK^4$.}
\label{cvrtable}
\end{table*}

\begin{table*}
\begin{center}
\begin{tabular}{llcll}
Method & COBE dataset & Q? & Result & Reference \\
\tableline
Correlation function & 1 year 53$\times$90 & N &
$n_{app} = 1.15^{+0.45}_{-0.65}$ & Smoot \etal\ (1992) \\
COBE:$\sigma_8$      & 1 year 53+90        & N &
$n_{pri} = 1 \pm 0.23 $ & Wright \etal\ (1992) \\
Genus \vs\ smoothing & 1 year 53           & Y &
$n_{app} = 1.7^{+1.3}_{-1.1}$ & Smoot \etal\ (1994)  \\
RMS \vs\ smoothing   & 1 year 53           & Y &
$n_{app} = 1.7^{+0.3}_{-0.6}$ & Smoot \etal\ (1994)  \\
Correlation function & 2 year 53$\times$90 & Y &
$n_{app} = 1.59^{+0.49}_{-0.55}$ & Bennett \etal\ (1994) \\
Correlation function & 2 year 53$\times$90 & N &
$n_{app} = 1.21^{+0.60}_{-0.55}$ & Bennett \etal\ (1994) \\
COBE\,:\,$1^\circ$ scale & 2 year NG           & N &
$n_{pri} = 1.15\pm 0.2$ & this paper \\
Cross power spectrum & 1 year (53A+90A)$\times$(53B+90B) & N &
$n_{app} = 1.69^{+0.45}_{-0.52} $  & this paper \\
Cross power spectrum & 2 year 53A$\times$53B & N &
$n_{app} = 1.41^{+0.75}_{-1.17} $  & this paper \\
Cross power spectrum & 2 year 53$\times$90 & N &
$n_{app} = 1.22^{+0.42}_{-0.46} $  & this paper \\
Cross power spectrum & 2 year (53A+90A)$\times$(53B+90B) & N &
$n_{app} = 1.32^{+0.39}_{-0.45} $  & this paper \\
Cross power spectrum & 2 year (53A+90B)$\times$(53B+90A) & N &
$n_{app} = 1.20^{+0.42}_{-0.46} $  & this paper \\
Cross power spectrum & 2 year NGA$\times$NGB & N &
$n_{app} = 1.41^{+0.75}_{-1.17} $  & this paper \\
Orthonormal functions & 2 year 53+90 & N &
$n_{app} = 1.02 \pm 0.4 $ & G\'orski \etal\ (1994) \\
\hline
\end{tabular}
\end{center}
\caption{Spectral index determinations}
\label{ntable}
\end{table*}

\clearpage

\ifms
\begin{center}
FIGURE CAPTIONS
\end{center}
\fi

\end{document}